# Directed Acyclic Graph Based Blockchain Systems

| Anand Devarajan | Erkan Karabulut |
|---|---|
| Technical University of Munich | Technical University of Munich |
| Germany | Germany |
| anand.devarajan@tum.de | erkan.karabulut@tum.de |

## ABSTRACT

**Blockchain technology has been revolutionizing many fields since last decade. It's true potential is not practically utilized yet. In a very short period of time, it has evolved twice - Smart contracts and Directed Acyclic Graph (DAG). DAG based blockchains currently referred to as Blockchain 3.0 solves many issues in the current conventional blockchain technologies including transaction fees, transaction approval times and scalability. In this paper, we present a comparative analysis of blockchain implementations based on DAG including IOTA, NxT, Byteball, Nano, DAGCoin, Fantom, XDAG and Caixapay. We discuss limitations of both conventional and DAG based blockchains and suggest when to prefer DAG based blockchains.**

## 1 INTRODUCTION

In the last decade, blockchain became one of the most researched technologies in the world. It has been discussed by scientists from different fields including Computer Science, Mathematics, Engineering, Decision Sciences, Medicine and Biochemistry[1].

Initially, it was designed just as a new form of currency which solves double-spending problems without a need for an authoritative third party[3]. Second big step of this new technology was smart contracts which allowed custom programming on the network[2]. This was again a revolutionary movement and it caused lots of other blockchain networks to be born.

Up until this point, there were two main building blocks, transactions and blocks. A block contains some limited amount of transactions. Each block on the network is connected to the previous block. When a new block is created, it directly approves transactions of previous blocks by simply including their hash value in itself.

By the time, it turned out that block structure itself can be also a security leak and it caused lots of scalability issues. This gave birth to Directed Acyclic Graph (DAG) based blockchain technologies which people call as "Blockchain 3.0".

As described in *section 2*, DAG based blockchains don't use a plain linear chain structure. Transactions are directly concatenated to each other asynchronously. Each of them approves an older transaction by including their hash value in itself. Removing block structure from the network also removed all the problems caused by blocks including long transaction approval times, high fee and scalability. These improvements make it possible to use blockchains for the applications which require a high amount of transactions such as IoT.

Scientific contribution of this paper includes:
- a comparative analysis of DAG based blockchains, including new implementations and comparison points in addition to previous study[5]
- defining limitations of DAG based blockchains
- providing a suggestion *(see Appendix A)* for when to prefer DAG based blockchains

The remaining has been structured as follows: *Section 2* points out downsides of conventional blockchains. *Section 3* elaborates how a DAG based blockchain works. *Section 4* consists of analysis of current popular DAG based blockchain implementations and a comparison table. In *Section 5*, we described limitations of DAG based blockchains. At last, *section 6* concludes the paper.

## 2 LIMITATIONS OF CONVENTIONAL BLOCKCHAIN TECHNOLOGY

In this section we will discuss the main limitations of conventional blockchain technology that gave birth to DAG based blockchains.

### 2.1 Scalability

In a conventional blockchain, after creation of a block it needs to be approved by the majority of the network which is inefficient. To tackle scalability issues researchers mostly work on two different approaches; tuning parameters and designing a new scalable architecture[6]. Both approaches bring some performance gain to the system. But they are still quite away from current payment systems. VISA handles ~65000 TPS[7] while Bitcoin can handle 7 TPS[8] and Ethereum can handle 15 TPS[9]. On the other hand, most of the DAG based systems don't use any block structure and in order to make a new transaction, each node approves 1 or more transactions in the system, allowing almost instant transactions. For instance, DAGCoin allows transactions in a few minutes and it is able to handle over 10,000 TPS[10].

## 2.2 Transaction Fee

In conventional blockchains, computational power of other nodes are required, e.g. PoW and PoS. This directly increases the total cost to make a transaction. Microtransactions are almost impossible as transaction fees could be more than intended transfer amount. In a DAG based system, involvement of other peers during a transaction isn't necessary. This feature of DAG system reduces transaction fees to almost zero and makes it possible to have microtransactions. Currently, the necessary fee for sending 10.000 USD with Bitcoin is 3.35 USD while sending it with DAGCoin is around 0.0005 USD[11].

## 2.3 Transaction Approval Time

In the PoW based systems, miners are required to do heavy, complex computations to approve new transactions. It can only be valid after each of those peers approve it. In the PoS systems, approval of a transaction depends on how busy the network is and how much the transaction initiator pays for fee. Eventually, this makes the approval process longer. On the contrary, finality in most of the DAG based system is usually in the magnitude of seconds as observed in *section 4*.

## 2.4 Dark Side of Decentralization

By its existence, blockchain systems are a chain of trust. Making a transaction means directly trusting the peers on the network on successful execution of the transaction. As described in Pervez et. al.'s work [5], decentralization can also cause adverse effects on the network. Malicious peers may try to play with transaction contents, prioritize some other transactions or drop them randomly. Decentralization also increases attack surface[4].

## 3 DAG BASED BLOCKCHAINS

A Directed Acyclic Graph is a type of directed graph that is structured in a way which doesn't allow passing through a node twice. This is what makes it acyclic. A directed tree is a good example of Directed Acyclic Graphs[12].

In DAG based blockchains, transactions are represented as nodes and approvements are represented as edges.

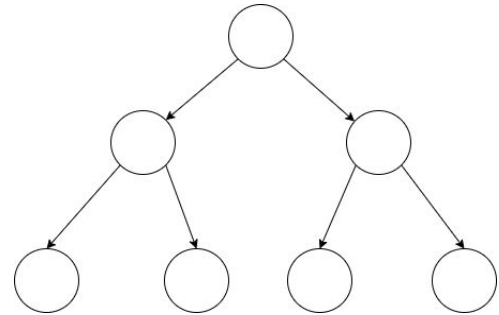

**Figure 1: Binary tree is an example of DAG**

When a new transaction occurs, other non-validated transactions are picked and attached to this new one by including it's hash value in itself. Therefore when there is an upcoming transaction, it is used to validate some previous transactions. There is no need to group transactions into blocks. This also speeds up the transaction approval process and reduces the amount of computational power required. Another important question here is - how do we choose the transactions to be validated? As an example, in IOTA, 2 transactions are randomly picked[13]. As all the transactions are referring to previous transactions, whenever a transaction is approved this also means all the transactions preceding it are also approved. Consequently the first transaction on the network which is called "genesis", is validated by all of the others. This also means that earlier transactions are much reliable then the current ones. Another issue that arises here is the consensus mechanism. Most common way to solve conflicts is to choose the chain of highly reliable transactions similar to conventional blockchains.

A DAG network can run asynchronously which may cause conflicting nodes to exist on the network at the same time. For instance, Consider X and Y nodes as conflicting nodes in Figure 2. Node X is validated by 3 nodes which are F, A and B while Y is validated by only C and D. Most common approach in this case is to pick X as valid and discard Y. On the other hand there is no unanimous way to treat discarded transactions. Each technology has their own way to handle these consensus-related problems.

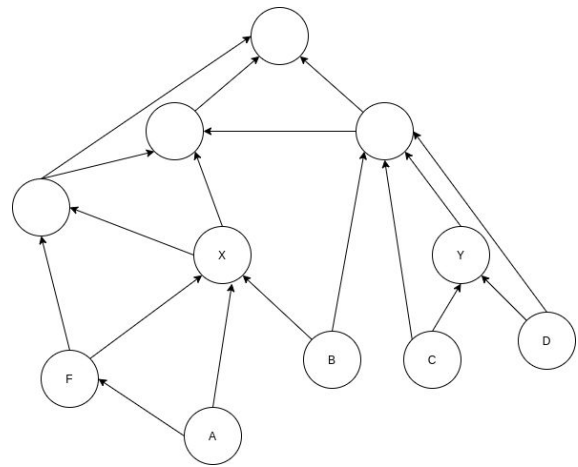

**Figure 2: A more complex DAG structure**

## 4 EXISTING DAG BASED BLOCKCHAINS

In this section, 8 of popular and unique DAG based blockchain implementations will be presented.

### 4.1 IOTA

IOTA is one of the popular DAG based cryptocurrency with the primary purpose of serving as default mode of payment in microtransactions among IoT devices[27]

**The tangle:** Tangle is the data structure behind IOTA. It is a type of directed graph which holds transactions. Transactions are represented as vertices of the directed graph. When a new transaction joins the graph, it must approve two previous transactions, thereby adding edges with them. There's always at least one not yet approved new transaction. These unapproved transactions are called tips. The selection strategy of these tips is to pick the previous transactions that form the crux of IOTA Tangle protocol[28]. All transactions together form the web of tangle.

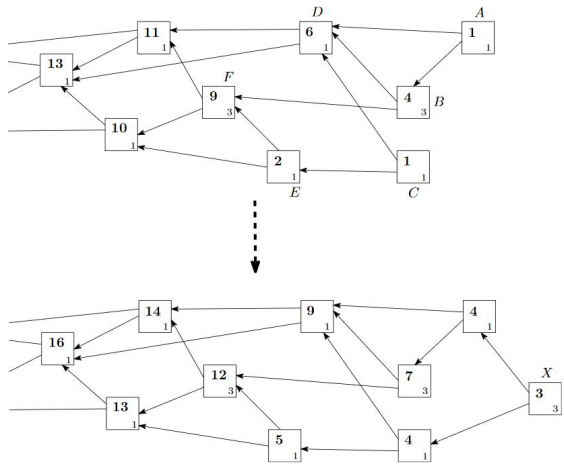

**Figure 3: A new transaction X joining tangle[13]**

**Consensus and MCMC:** IOTA assigns weight and confidence to each transaction for it to be pushed into consensus. The weight depends on the number of transactions that approves it. The number of transactions is determined by poisson point process and denoted by transaction rate λ. The weights to the transaction are allotted based on a special weighted random walk using "Markov Chain Monte Carlo (MCMC)" Algorithm[13]. This prevents the lazy tips that conveniently approves only the very early transactions from becoming a valid transaction. A certain bias parameter alpha is needed for assigning confidence to the transaction's legitimacy, resulting in the hybrid of unweighted and super-weighted walks along the tangle[29].

**The Coordinator:** IOTA's architecture requires a high rate of transaction for it to be secure & decentralized since low λ can result in chain formations leading to only one tip available for approval, meaning a group with huge compute resources can double spend by approving transactions convenient to them. To prevent confirmation confidence of these transactions from becoming 100%, IOTA introduces a mechanism called coordinator. Every two minutes, a milestone transaction is issued by the coordinator, and all transactions approved by it are considered to have a confirmation confidence of 100%, immediately. Using the coordinator, the double spent transaction will never be approved. This acts as a protective mechanism while the IOTA network keeps growing toward the necessary activity from adoption needed to make the network mature. At that point the IOTA Foundation plans to shut down the Coordinator and let tangle grow on its own[30].

### 4.2 NxT

NxT is one of the first blockchain technologies that uses a DAG based structure[14]. It provides tools for a decentralized financial platform[15]. It is 100% based on the Proof-of-Stake concept[16]. Main focus point of Nxt is to provide a fully decentralized peer-to-peer system for the electronic economy[17]. New blocks are generated every 60 seconds on average. It has its own unique algorithm for block creation which is referred to as forging.

#### 4.2.1 Forging

There are 3 different notions in block creation (forging). Base target value, target value and cumulative difficulty. Base target value is a threshold value which differs from block to block and it is equal to previous block's base target value multiplied by the amount of time in seconds that passed while creating that block.

Target value is specific to each account and it is a multiplication of base target value, time passed after last block creation in seconds and effective balance of the account. Therefore, as the account balance increases, the target value also increases. As the target value increases by time, even if there are a small number of accounts in the network, one of them eventually achieves to create the next block.

TABLE I: Comparison of Existing DAG Technology

| Feature/Technology | NxT | IOTA | DAGCoin | ByteBall | Nano | XDAG | Fantom | Caixapay |
|---|---|---|---|---|---|---|---|---|
| Average Confirmation Time | ~60 seconds | 2 milliseconds theoretically, ~60 seconds in current implementation | 30 seconds | 30 seconds | < 10 seconds | - | 1 - 2 seconds | < 1 second |
| Spam Protection | Pos during Forging | PoW | Re-hashing of previous blocks + fee | Re-hashing of previous blocks + fee | PoW | None. only security algorithms are present | Fees | Confirming previous blocks |
| Open Source/Decentralized | -/Decentralized | Coordinator is close source / Partially | -/Decentralized | Fully / Decentralized | Fully / Decentralized; Delegation is present | Fully / Decentralized | Fully / Decentralized | -/Decentralized |
| State of the Network | Advanced | Advanced | Advanced | Advanced | Advanced | No data | Advanced | Advanced |
| Fees | min of 1 NxT(~1 cent) | zero | 0.0005 DAGCoin | 1 MB storage fee is 0.033 $ | zero | - | very low fee in form of FTG | Almost 0 |
| Main Selling Points | Asset exchange, voting system, data cloud, marketplace, monetary system, coin shuffling…([https://www.jelurida.com/nxt/nxt-features](https://www.jelurida.com/nxt/nxt-features)) | IoT Microtransactions and Data transfer | Decentralized immutable storage of data, near-zero fee and instant transactions | Atomic exchange, multi signature, immutable storage, regulated assets, on-chain oracles, settlement finality | Microtransactions - value transfer. | First mineable DAG based token with privacy features. | Smart contracts infrastructure for transactions and data sharing, interoperable with Ethereum, DeFi. | Instant anonymous transactions with almost 0 transaction fee |
| Market Cap | ~ $10 mln | ~ $670 mln | 1,189,500,000 € | ~ $19 mln | ~ $132 mln | No data | ~ $25.1 mln | 158,354 $ |
| Coin Supply | 1 bln NXT | 2,779,530,283 MIOTA | 9 bln DAGCoin | 1,000,000 GBYTE | 133,248,297 NANO | 1,070,972,160 XDAG | 2,132,239,133 FTM | 131,500,000 CXP |
| New Incoming Big Feature | - | Coordicide, Bee nodes, Smart Contracts | - | - | New PoW spam protection algorithm | - | Custom smart contract engine - FVM | - |
| Smart Contract | - | WIP | - | Yes | No | No | Yes | - |
| Consensus Algorithm | Cumulative difficulty of previous blocks | FPC and MCMC | Longest trustable chain | Longest trustable chain | Open Representative Voting | Longest trustable chain | Lachesis - an aBFT algorithm | Longest trustable chain |
| Conditional Transaction | - | No | - | Yes | No | No | Yes | - |

$$T = T_b \times S \times B_e$$

$T$ : *New target value*
$T_b$ : *Base target value*
$S$ : *Time since the last block in seconds*
$B_e$ : *Effective balance of the account*

At this point, another problem raises. Since all the nodes can query the accounts and see their balance, they will be able to make a quite precise guess on which account will generate the next block. As a response to this, forging is only possible when 1440 blocks are generated, making prediction quite harder.

#### 4.2.2 Proof of Stake Model of Nxt

In the Nxt blockchain, new tokens are not created as a result of forging. Reward of a block creation is the sum of the transaction fees in a block. So this is a redistribution process called forging and not mining. In case of an ambiguity, Nxt chooses the block or chain fragment with the highest cumulative difficulty. So the cumulative difficulty is used as a consensus mechanism.

$$D_{cb} = D_{pb} + \frac{2^{64}}{T_b}$$

$D_{cb}$ : *Difficulty of the current block*
$D_{pb}$ : *Difficulty of the previous block*
$T_b$ : *Base target value for the current block*

Each block is generated based on verifiable, unique and statistically almost unpredictable information gathered from previous blocks and accounts itself. Therefore each block is connected to each other and in any time it can be traced back to the genesis block. Since the time passed during the block generation process is also a parameter of the block generation process itself, it is possible to predict a time for new block generations which is 60 seconds by default. In order to prevent an attacker from creating another chain starting from genesis block, Nxt allows chain re-organization only up to 720 blocks behind the current block. After 10 blocks are added to the chain, all the blocks before them are accepted as safe.

### 4.3 ByteBall

Byteball (currently rebranded as Obyte) is a decentralized tamper proof public data storage. It allows to store arbitrary data including data that represents transferrable values such as currencies, property titles, debts, shares[19] It provides atomic exchange, multi signature, immutable storage, regulated assets, on-chain oracles and settlement finality features[18]. Each storage unit on the Byteball network includes one or more hashes of the previous units which helps to verify previous units and establish their partial order. Therefore there are no blocks and no mining process.

Everyone is allowed to add new storage units after signing them and paying a necessary fee which is directly proportional to its size. This fee goes to the later storage units that verifies them by including their hash in its own. Any type of assets can be issued on the ByteBall network along with the rules that govern their transferability. ByteBall has an internal currency called bytes. All the assets on the network including bytes can be sent between peers on the network. In case of a double-spending both transactions are stored in the database but only the earlier one is accepted as valid. Overall order is established by choosing a single chain which is signed by known users called witnesses.

### 4.4 Fantom

FANTOM is a DAG based smart contract platform [31]. It uses a consensus algorithm called Lachesis, an aBFT consensus algorithm that claims to solve the classical blockchain trilemma[31]. The architecture of FANTOM uses modular architecture – OPERA and XAR being the two layers. It is interoperable and compatible with Ethereum smart contracts by using its bridge layer OPERA[32]. XAR network is the framework for use-cases like DeFi. Any node can involve consensus and are called Lachesis Node[33]. Each node stores a local DAG composed of events that contains transactions within them. DAG is enforced to establish a before-after relationship locally and independently within the node. New events are used to vote for the events in 2-3+ previous virtual elections simultaneously [34]. Unlike PoW or PoS, no confirmed blocks are sent among nodes but instead only the events are exchanged periodically and observed.

### 4.5 Nano

The original Nano (formerly RaiBlocks) is one of the first Directed Acyclic Graph (DAG) based cryptocurrencies [35]. Nano is a platform with low-latency payment that wants minimum resources, making it ideal for P2P transactions. It uses a novel block-lattice architecture, where each node has its own blockchain and only the owner can update this account-chain. The transaction updates are asynchronous. Nano uses a consensus mechanism called ORV[36], where every account chooses their representative, which themselves chose their principal representative that engages in achieving quorum. At the time of this writing, 50.4 million transactions have been processed by the Nano network, yielding a total blockchain size of mere 25.5 GB[37].

### 4.6 XDAG

XDAG is a DAG based mineable cryptocurrency that aims to create a decentralized payment system[38]. Each block is considered an address and contains only one transaction. Main chain is chosen according to the one that has maximum difficulty. New coins are created about every 64 seconds[39]. Valid transactions are strictly ordered depending on main chain and links order. Double spending is prohibited because only the first concurrent transaction is approved.

### 4.7 CaixaPay

The purpose of CaixaPay is to create a DAG based network which allows instant transactions and almost zero transaction fees[20]. Therefore it is designed to be used in everyday situations. It is an anonymous cryptocurrency which means transactions on the network cannot be linked to a particular user or real world identity. The consensus algorithm of CaixaPay is to choose a main chain which gravitates towards units issued by commonly recognised reputable users, so called witnesses[21].

Each incoming transaction confirms at least one previous transaction. Transactions don't come together to create a block and therefore each transaction doesn't have to wait for a block creation. CaixaPay gets benefit from these features of a DAG enabling almost no cost, instant microtransactions[22].

### 4.8 DAGCoin

DAGCoin is a public decentralized immutable storage of arbitrary data, including data of social value such as money[23]. DAGCoin has near-zero transaction fees, around 0.0005 DAGcoins, 30 seconds of transaction time on average and freedom to transact money to anyone all over the world without any limitation[24].

It was initially built on top of Byteball as an asset on the network. Currently, DAGCoin has it's own codebase. The consensus mechanism is basically to choose the chain which is issued by commonly recognized reputable users called witnesses.

## 5 CURRENT LIMITATIONS OF DAG BASED BLOCKCHAINS:

While DAG scales really well with higher number of transactions, In platforms like IOTA - the network is highly susceptible to security vulnerabilities such as large weight attack when the transaction rate is low[25]. Also most DAG based blockchain systems are still in prototype stage and yet to be tested by the mass. For instance, A highly critical vulnerability was found by researchers leveraging IOTA's home-built hash function Curl-P-27[26] after the mainnet was released. Though it's fixed now, it shows that there might still be zero day vulnerabilities that are yet to be discovered. DAG based blockchains still can't provide generalized smart contracts and even though FTM provides smart contracts, it still incurs the cost in performance overhead of executing it in EVM. There are also some centralized elements like 'coordinator' in IOTA[30] that ought to be removed to make them fully decentralized.

## 6 CONCLUSION AND FUTURE WORKS

DAG mainly solves one of the key weaknesses of blockchain - scalability, making it practically viable in use cases such as IoT and microtransactions. Both DAG and conventional blockchain attempts to solve the same Byzantine Generals problem. From our analysis (*see Appendix A*) and comparisons in this paper, we suggest to prefer DAG over conventional permissionless blockchain under two main conditions:

- If scalability is a top priority, in addition to decentralization.
- If a high number of transactions are assured at any given time.

Enabling smart contracts in DAG based blockchains can give a huge boost to its adoption, ranging from payment automation to problems in logistics. Most importantly, There is a need to design and develop robust algorithms and systems to preserve safety of the DAG network during time periods when transaction rate is low and also there is a need for optimal transaction validation algorithms as well.

# APPENDIX A

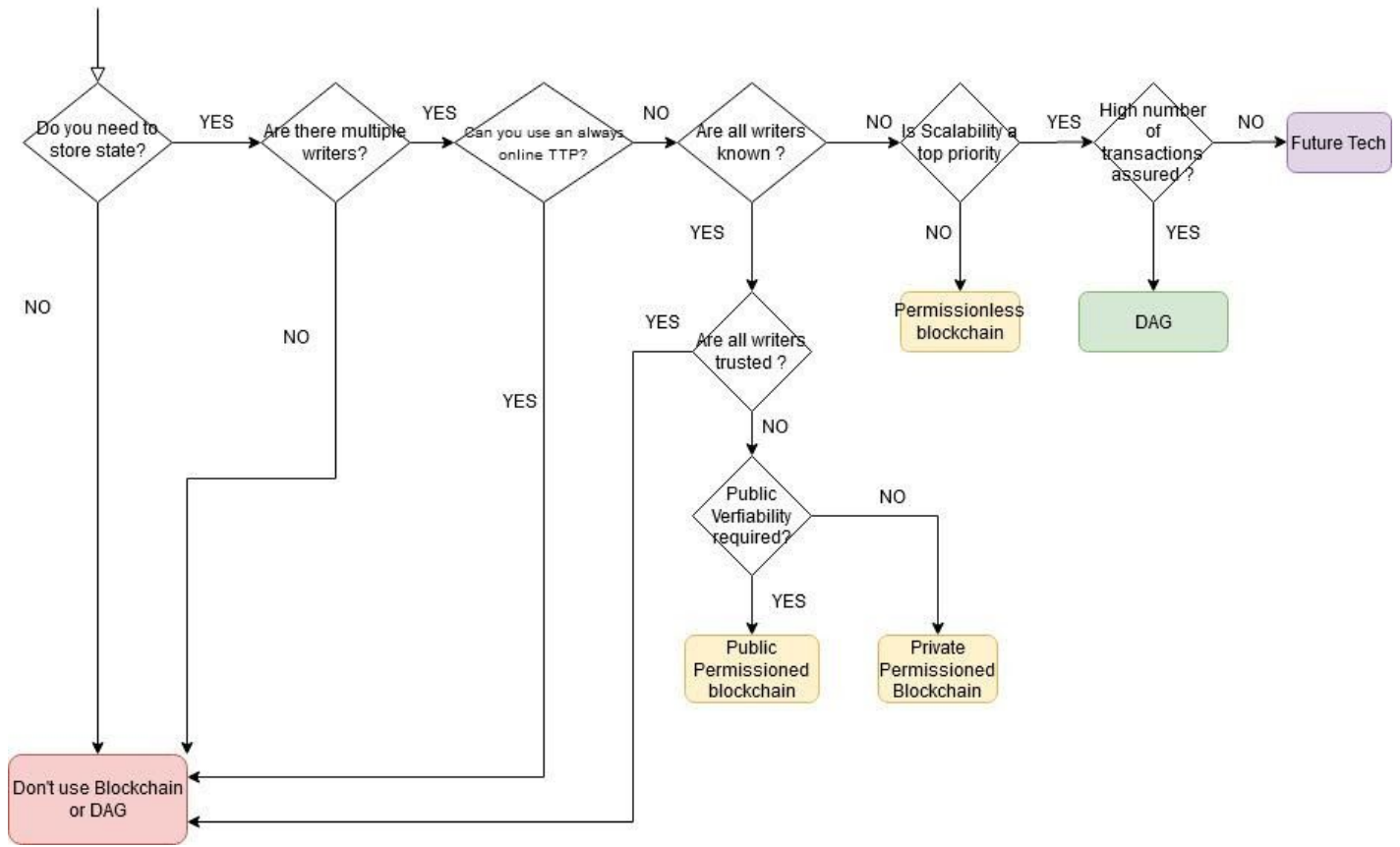

**TTP** -- Trusted Third Party/Parties

**Future Tech** -- Cutting edge DLT that will solve the classical blockchain trilemma in future. Some examples of current contestants for this place include Hashgraph and Radix. Both are architecturally and substantially different DLT from blockchain and DAG.